\documentclass[prl,twocolumn,showpacs,preprintnumbers,amsmath,amssymb,citeautoscript]{revtex4-1}

\usepackage{graphicx}
\usepackage{amssymb, amsfonts, amsmath}
\usepackage{dcolumn}
\usepackage{bm}
\usepackage{color}

\newlength{\figwidth}
\begin{document}


\title{Controllable Rashba spin-orbit interaction in artificially engineered superlattices involving the heavy-fermion superconductor CeCoIn$_5$}

\author{M.~Shimozawa$^1$}
\author{S.~K.~Goh$^{1,2}$}
\author{R.~Endo$^1$}
\author{R.~Kobayashi$^1$}
\author{T.~Watashige$^1$}
\author{Y.~Mizukami$^1$}
\author{H.~Ikeda$^1$}
\author{H.~Shishido$^3$}
\author{Y.~Yanase$^4$}
\author{T.~Terashima$^5$}
\author{T.~Shibauchi$^1$}
\author{Y.~Matsuda$^{1}$}

\affiliation{$^1$Department of Physics, Kyoto University, Kyoto 606-8502, Japan}
\affiliation{$^2$Cavendish Laboratory, University of Cambridge, JJ Thomson Avenue, Cambridge CB3 0HE, United Kingdom}
\affiliation{$^3$Department of Physics and Electronics, Osaka Prefecture University, Osaka 599-8531, Japan}
\affiliation{$^4$Department of Physics, Niigata University, Niigata 950-2181, Japan}
\affiliation{$^5$Research Center for Low Temperature and Materials Science, Kyoto University, Kyoto  606-8501, Japan}

\date{\today}

\begin{abstract}
By using a molecular beam epitaxy technique, we fabricate a new type of superconducting superlattices with controlled atomic layer thicknesses of alternating 
blocks between heavy fermion superconductor CeCoIn$_5$, which exhibits a strong Pauli pair-breaking effect, and nonmagnetic metal YbCoIn$_5$. The introduction of the thickness modulation of YbCoIn$_5$ block layers breaks the inversion symmetry centered at the superconducting block of CeCoIn$_5$. This configuration leads to dramatic changes in the temperature and angular dependence of the upper critical field, which can be understood by considering the effect of the Rashba spin-orbit interaction arising from the inversion symmetry breaking and the associated weakening of the Pauli pair-breaking effect. Since the degree of thickness modulation is a design feature of this type of superlattices, the Rashba interaction and the nature of pair-breaking are largely tunable in these modulated superlattices with strong spin-orbit coupling. 


\end{abstract}


\maketitle

Among the existing condensed matter systems, the metallic state with the strongest electron correlation effects is achieved in heavy fermion materials with 4$f$ or 5$f$ electrons.   In these systems, a very narrow conduction band is formed at low temperatures through the Kondo effect.  In particular, in Ce(4$f$)-based compounds,  strong electron correlations within the narrow band strikingly enhance the quasiparticle effective mass.  As a result of notable many-body effects, a plethora of fascinating physical phenomena including unconventional superconductivity with non-$s$-wave pairing symmetry appears {\cite{Pfl09}}. The unconventional pairing symmetry and the associated exotic superconducting properties have mystified researchers over the past quarter century.
 
Recently, it has been suggested that the inversion symmetry breaking (ISB) together with strong spin-orbit interaction can dramatically affect the superconductivity,  giving rise to a number of novel phenomena such as  anomalous magneto-electric effects {\cite{Lu08}} and topological superconducting states {\cite{Tan09,Sat09,Qi11}}.   It has also been pointed out  that such phenomena are more pronounced  in strongly correlated electron systems {\cite{Fuj07}}.  The inversion symmetry imposes important constraints on the pairing states: In the presence of  inversion symmetry, Cooper pairs are classified into a spin-singlet or triplet state,  whereas in the absence of inversion symmetry, an asymmetric potential gradient $\nabla V$ yields a spin-orbit interaction that breaks parity, and  the admixture of spin singlet  and  triplet states is possible {\cite{Gor01,Fri04}}.   For instance, asymmetry of the potential in the direction perpendicular to the two-dimensional (2D) plane $\nabla V \parallel [001]$ induces Rashba spin-orbit interaction $\alpha_R {\bm g}({\bm k})\cdot{\bm \sigma}\propto({\bm k} \times \nabla V)\cdot {\bm \sigma}$, where ${\bm g}({\bm k})=(-k_y,k_x,0)/k_F$, $k_F$ is the Fermi wave number, and ${\bm \sigma}$ is the Pauli matrix.  Rashba interaction splits the Fermi surface into two sheets with different spin structures: the energy splitting is given by $\alpha_R$, and the spin direction is tilted into the plane, rotating clockwise on one sheet and anticlockwise on the other.  When the Rashba splitting exceeds the superconducting gap energy ($\alpha_R > \Delta$), the superconducting properties are dramatically modified.  Therefore,  in Ce-based superconductors, where the spin-orbit interaction is generally significant, the introduction of ISB makes the systems a fertile ground for observing exotic  properties.  Although there are bulk heavy fermion superconductors with ISB such as CePt$_3$Si {\cite{Bau04}} and CeRhSi$_3$ {\cite{Kim05}},  their superconductivity often coexists with magnetic order and the degree of the ISB is hard to be controlled.  Thus the systematic influence of ISB on unconventional superconductivity remains an open question.

CeCoIn$_5$ is a  heavy fermion superconductor {\cite{Pet01}} which hosts a wide range of fascinating superconducting properties including extremely strong Pauli pair-breaking effect {\cite{Iza01,Tay02,Bia02,Oka08,Bia08}} and an associated possible Fulde-Ferrell-Larkin-Ovchinnikov (FFLO) state with a novel pairing state ({\boldmath $k$}$\uparrow$, {\boldmath $-k+q$}$\downarrow$) {\cite{Rad03,Bia03,Kak05,Mat07,Kum11}}. Although the crystal structure of bulk CeCoIn$_5$ possesses the inversion symmetry,  band structure calculations suggest that even a small degree of ISB can induce a large Rashba splitting of the Fermi surface {\cite{Ikeda}}.   Recently, a state-of-the-art molecular beam epitaxy technique has been developed to fabricate the $c$ axis oriented artificial superlattices with alternating layers of  CeCoIn$_5$  and  nonmagnetic, nonsuperconducting metal YbCoIn$_5$ with controlled atomic layer thicknesses {\cite{Shi10,Miz11,Goh12,Shim}}.   In  these superlattices, RKKY interaction between the Ce atoms in neighboring CeCoIn$_5$ block layers is substantially reduced {\cite{Pet13}}.  Moreover, the superconducting proximity effect  between CeCoIn$_5$- and YbCoIn$_5$- layers is  negligibly small due to  the large Fermi velocity mismatch {\cite{She12}}.  In fact, it has been shown that in the superlattices with 4--6 unit cell thick CeCoIn$_5$ layers, whose thickness  is comparable to the perpendicular coherence length $\xi_{\perp}\sim$3--4\;nm,  2D heavy fermion superconductivity is realized {\cite{Miz11}}.   In these superlattices,  the importance of the {\it local} ISB at the interface between  CeCoIn$_5$ and YbCoIn$_5$ 
has been emphasized experimentally through the peculiar angular variation of upper critical field $H_{c2}$, which can be interpreted as a strong suppression of the Pauli pair-breaking effect {\cite{Goh12}}.  
Theoretical studies also suggest that when the interlayer hopping integral is comparable to or smaller than the Rashba splitting $(t_c\alt \alpha_R)$, the local ISB plays an important role in determining the nature of the superconducting state {\cite{Mar12}}. This appears to be the case for the CeCoIn$_5$/YbCoIn$_5$ superlattices. 

To advance the understanding of the effect of the ISB on the superconducting ground state, we designed and fabricated a new type of superlattices, i.e. \textit{modulated} superlattices in which the thickness of CeCoIn$_5$ is kept to $n$ for the entire superlattice, while the thickness of YbCoIn$_5$ alternates between $m$ and $m'$ from one block layer to the next, forming a ($n$:$m$:$n$:$m'$) $c$ axis oriented superlattice structure. We demonstrate that, through the introduction of the thickness modulation of YbCoIn$_5$ layers, the Rashba effect in each superconducting  CeCoIn$_5$ block layer  is largely tunable, leading to profound changes in the nature of superconductivity.  This ``block tuning" of Rashba interaction appears to  pave the way for obtaining novel superconducting states.

We study superlattices both without and with thickness modulation of YbCoIn$_5$ layers, $m=m'$ (Fig.~1(a)) and $m \neq m'$ (Fig.~1(b)), respectively (for the fabrication method, see Sec.\,S1 in {\cite{SI}}).  We denote the former as S-type and the latter as S$^{\ast}$-type superlattices. In both superlattices,  the asymmetric potential gradient associated with the  local ISB, $-\nabla V_{\rm local}$, gives rise to the Rashba splitting. This splitting is the largest at the top and bottom CeCoIn$_5$ layers and vanishes at the middle layer, as shown by the  green (small) arrows in Figs.~1(a) and (b).    The critical difference between the S- and S$^{\ast}$-type superlattices is that, as  illustrated in Figs.\;1(a) and (b),  for the S-type superlattices, the middle Ce plane in a given CeCoIn$_5$ block layer is a mirror plane, whereas for the S$^{\ast}$-type it is not.  In the S$^{\ast}$-type superlattices, therefore, the additional ISB along the $c$-axis  can be introduced to the superconducting CeCoIn$_5$ block layers.  The asymmetric potential gradient associated with the YbCoIn$_5$  thickness modulation, $-\nabla V_{\rm block}$, point to the opposite direction in the neighboring CeCoIn$_5$-block layers as shown by the orange (large) arrows in Fig.~1(b).  (We note that even in the S$^{\ast}$-type one can find mirror planes in YbCoIn$_5$ layers but here we focus mainly on the ISB in the superconducting planes.)  We expect that the degree of this ``{\it block layer} ISB" (BLISB) can be enhanced with increasing $|m-m'|$, which represents the degree of thickness modulation.  

Here we stress that one of the most remarkable effects of the ISB on the superconductivity appears in the magnetic response: the Zeeman term in magnetic field in the presence of Rashba interaction is given by $\pm {\bm g}({\bm k})\cdot \mu_0{\bm H}$, which leads to a strong suppression of the Pauli pair-breaking effect, in particular for  ${\bm H}\parallel c $ where ${\bm g}({\bm k})$ is always perpendicular to ${\bm H}$ {\cite{Gor01,Fri04}}.   It is therefore of tremendous interest to experimentally introduce the ISB in CeCoIn$_5$ with strong Pauli-limited superconductivity, for this will allow the study of its interplay with the strong Pauli pair-breaking effect and its influence on the unconventional superconductivity of an archetypal heavy fermion superconductor.

\begin{figure}[t]
\begin{center}
\includegraphics[width=0.9\linewidth]{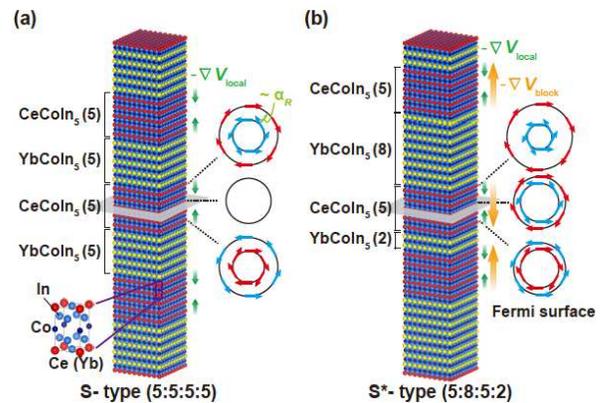}
\caption{(color online). Schematic representations of the CeCoIn$_5$$(n)$/YbCoIn$_5$$(m)$/CeCoIn$_5$$(n)$/YbCoIn$_5$$(m')$ artificial superlattices. (a) S-type ($m=m'$): Superlattice with alternating layers of 5-UCT CeCoIn$_5$ and 5-UCT YbCoIn$_5$, ($n$:$m$:$n$:$m'$) = (5:5:5:5).  The middle CeCoIn$_5$ layer in a given CeCoIn$_5$ block layer indicated by the gray plane is a mirror plane.  The green (small) arrows represent the asymmetric potential gradient associated with the local ISB, $-\nabla V_{\rm local}$.  The Rashba splitting occurs at the interface between the CeCoIn$_5$ and YbCoIn$_5$ 
 due to the local ISB. The spin direction is rotated in the $ab$ plane and is opposite between the top and bottom CeCoIn$_5$ layers.  (b) S$^{\ast}$-type ($m\ne m'$):  5-UCT CeCoIn$_5$ block layers are sandwiched by 8- and 2-UCT YbCoIn$_5$ layers, ($n$:$m$:$n$:$m'$) = (5:8:5:2).  The middle CeCoIn$_5$ layer (gray plane) is not a mirror plane.  The orange (large) arrows represent the asymmetric potential gradient associated with the YbCoIn$_5$ layer thickness modulation $-\nabla V_{\rm block}$.}
\end{center}
\vspace{-5mm}
\end{figure}

\begin{figure}[t]
\begin{center}
\includegraphics[width=0.95\linewidth]{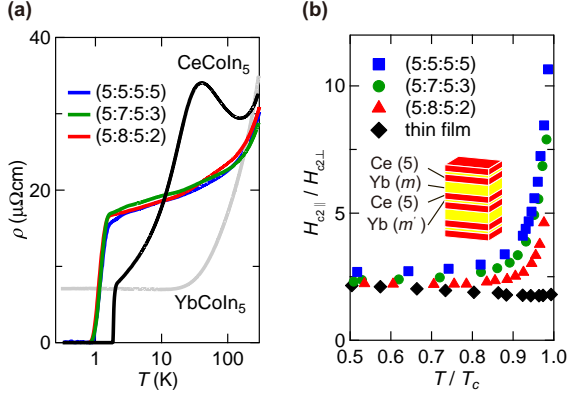}
\vspace{-2mm}
\caption{(color online). (a)  Temperature dependence of the resistivity in each superlattice, along with that of 120-nm-thick CeCoIn$_5$ and YbCoIn$_5$ epitaxial thin films. $T_c$, defined as the temperature where the resistance is 50\% of the normal state resistance, is 1.16, 1.18 and 1.13\,K for ($n$:$m$:$n$:$m'$) = (5:5:5:5), (5:7:5:3) and (5:8:5:2), respectively.  (b) Temperature dependence of the anisotropy of upper critical field $H_{c2}$, $H_{c2\parallel}/H_{c2\perp}$  for the superlattices and CeCoIn$_5$ thin film.
}
\end{center}
\vspace{-5mm}
\end{figure}

To examine how the BLISB affects the superconductivity of the 2D CeCoIn$_5$ block layers, we fabricated the S- and S$^{\ast}$-type superlattices, which consist of 5 unit-cell thick (UCT)  CeCoIn$_5$ ($n=5$) sandwiched by $m$- and $m'$-UCT YbCoIn$_5$, whose total number is fixed as $m+m'=10$ (For the structural characterization, see Sec.\,S2 in Ref.\,\cite{SI}).   As shown in Fig.\;2(a), the resistivity of the S-type superlattice with ($n$:$m$:$n$:$m'$) = (5:5:5:5) and S$^{\ast}$-type superlattices with ($n$:$m$:$n$:$m'$) = (5:7:5:3) and (5:8:5:2) exhibits very similar behavior. Figure\;2(b) shows the anisotropy of upper critical fields $H_{c2\parallel}/H_{c2\perp}$.  In sharp contrast to the CeCoIn$_5$ thin film with thickness of 120\,nm, $H_{c2\parallel}/H_{c2\perp}$ of superlattices exhibits a diverging behavior while approaching $T_c$, indicating a 2D superconducting behavior.

Figure\;3 displays the temperature dependence of $H_{c2\perp}$ normalized by the orbital limited upper critical field without the Pauli pair-breaking effect at $T$ = 0\;K, $H_{c2\perp}^{\rm orb}(0)$,  for ($n$:$m$:$n$:$m'$) = (5:5:5:5), (5:7:5:3) and (5:8:5:2).  Here $H_{c2\perp}^{\rm orb}(0)$ is calculated by the  Werthamer-Helfand-Hohenberg (WHH) formula {\cite{WHH66}},  $H_{c2\perp}^{\rm orb}(0)=-0.69T_c({\rm d}H_{c2\perp}/{\rm d}T)_{T_c}$.  We also include the two extreme cases, $H_{c2\perp}/H_{c2\perp}^{\rm orb}(0)$ for the bulk CeCoIn$_5$ {\cite{Tay02}}, in which superconductivity is dominated by Pauli paramagnetism, and WHH curve without the Pauli pair-breaking effect.  What is remarkable is that  at low temperatures $H_{c2\perp}/H_{c2\perp}^{\rm orb}(0)$ is enhanced with increasing $|m-m'|$. 
This is in sharp contrast to the case of bulk Ce$_{1-x}$Yb$_x$CoIn$_5$ system {\cite{Cap10}},  where the chemical substitution of Ce with Yb does not change the whole temperature dependence of $H_{c2\perp}/H_{c2\perp}^{\rm orb}(0)$. Because the resistivity of these superlattices shows a very similar temperature dependence with similar $T_c$ as displayed in Fig.\;2(a), this enhancement cannot be attributed to the difference in the electron scattering rate or impurity concentration. We can also rule out the possibility that the enhancement is related to the difference in the spin-orbit scattering (see Sec.\;S3 in {\cite{SI}}). The enhancement is rather associated with the reduction in the Maki parameter $\alpha_M$, which represents the ratio of the orbital-limited upper critical field to the Pauli-limited one. 
Thus the present results suggest that the BLISB modifies the superconducting properties, leading to the relative suppression of the Pauli pair-breaking effect with respect to the orbital pair-breaking effect.

\begin{figure}[t]
\begin{center}
\includegraphics[width=0.8\linewidth]{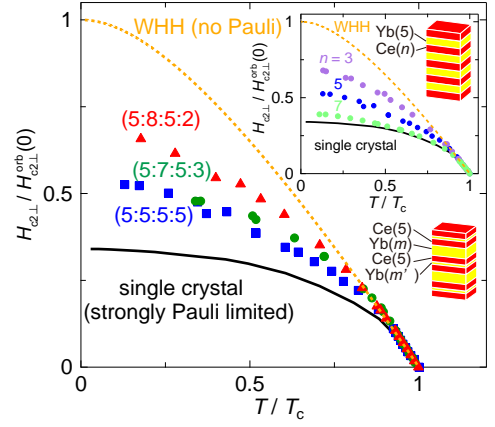}
\end{center}
\vspace{-9mm}
\caption{
(color online). Normalized upper critical field $H_{c2\perp}/H_{c2\perp}^{\rm orb}(0)$ as a function of $T/T_c$ for S$^\ast$-type superlattices with ($n$:$m$:$n$:$m'$) = (5:7:5:3) and (5:8:5:2) compared with that for S-type (5:5:5:5).  We also plot $H_{c2\perp}/H_{c2\perp}^{\rm orb}(0)$ for CeCoIn$_5$ single crystal with strong Pauli pair-breaking effect and the WHH curve without the Pauli pair-breaking effect.   Inset: 
The same plot 
for S-type CeCoIn$_5$($n$)/YbCoIn$_5$(5) superlattices.}
\end{figure}

The suppression of the Pauli pair-breaking effect in the S$^{\ast}$-type superlattices can be demonstrated clearly by the angular dependence of $H_{c2}$. Figure\;4(a) shows $H_{c2}(\theta)$ of the (5:8:5:2) superlattice determined by the resistive transitions in magnetic fields, where $\theta$ is the angle between $\bm{H}$ and the $a$-axis. $H_{c2}(\theta)$ increases monotonically without saturation as $|\theta|$ decreases and a distinct cusp behavior is observed at $\theta=0^\circ$. In sharp contrast,  in (5:7:5:3) and (5:5:5:5) superlattices, the cusp structure is not observed and $H_{c2}(\theta)$ is smooth for all $\theta$ (Fig.\;4(b)). This remarkable difference of $H_{c2}(\theta)$ between S-type (5:5:5:5) and S$^{\ast}$-type (5:8:5:2) superlattices is highly unusual because the CeCoIn$_5$ block layers in each superlattice have the same thickness ($n=5$), and hence similar angular variation of $H_{c2}(\theta)$ is expected, in particular near $T_c$ where $\xi_{\perp}(T)$ well exceeds the thickness of CeCoIn$_5$ block layer. In order to be more quantitative,  we analyze the data using the model below {\cite{Goh12}}:
\begin{equation}
\left[\frac{H_{c2}(\theta)\cos\theta}{H_{c2}(0^{\circ})}\right]^2+\beta_{T}\left|\frac{H_{c2}(\theta)\sin\theta}{H_{c2}(90^{\circ})}\right|+\beta_{P}\left[\frac{H_{c2}(\theta)\sin\theta}{H_{c2}(90^{\circ})}\right]^2=1,
\end{equation}
where $\beta_{T}$ ($\geq$ 0) and $\beta_{P}$ ($\geq$ 0) are fitting parameters with  $\beta_{T}+\beta_{P}=1$.  In Eq.\;(1) $(\beta_{T},\beta_{P})=(1,0)$ represents the so-called Tinkham model {\cite{Tin96}} which describes $H_{c2}(\theta)$ in the 2D thin film with thickness smaller than $\xi_{\perp}$ in the absence of  the Pauli pair-breaking effect. In the Tinkham model, cusp appears at $\theta=0$ as a result of the vortex formation due to the orbital pair-breaking effect in a slightly tilted field, which strongly suppresses $H_{c2}$. On the other hand, $(\beta_{T},\beta_{P})=(0,1)$ represents the anisotropic model, which describes $H_{c2}(\theta)$ of 2D superconductors when Pauli pair-breaking effect dominates. In this case, cusp does not appear because $H_{c2}(\theta)$ is determined by the anisotropy of $g$-factor, which changes smoothly with $\theta$. Thus $\beta_{T}/\beta_{P}$ quantifies the relative importance of orbital and Pauli pair-breaking effects. With this model, an excellent description of $H_{c2}(\theta)$ is achieved (Fig.\;4(c)).   Figure\;4(d) shows $\beta_{T}/\beta_{P}$ for  several superlattices at fixed reduced temperatures.  For the S$^\ast$-type superlattices, when going from (5:7:5:3) to (5:8:5:2),  $\beta_{T}/\beta_{P}$ is strongly enhanced, indicating the suppression of the Pauli pair-breaking effect. This result is consistent with the low-temperature enhancement of $H_{c2\perp}/H_{c2\perp}^{\rm orb}(0)$ with increasing $|m-m'|$ shown in Fig.\;3.  Thus, both the enhancement of $H_{c2\perp}(T)/H_{c2\perp}^{\rm orb}(T=0)$ in perpendicular field and the angular variation of $H_{c2}(\theta)$ around the parallel field indicate that the ISB in the direction perpendicular to the $ab$-plane strongly affects the superconductivity through the suppression of Pauli paramagnetism, when $|m-m'|$ is tuned from 4 to 6. This result can be understood if the Rashba splitting begins to exceed the superconducting gap energy when $|m-m'|$ reaches a threshold value between 4 and 6.

\begin{figure}[t]
\begin{center}
\includegraphics[width=0.95\linewidth]{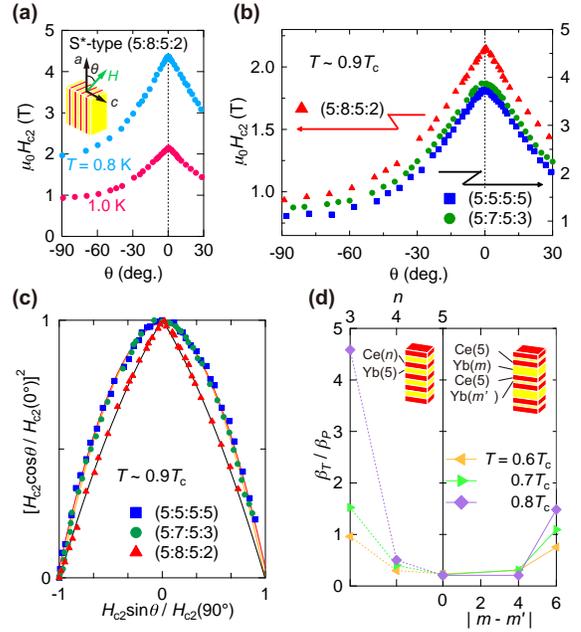}
\end{center}
\vspace{-10mm}
\caption{(color online). Angular dependence of $H_{c2}$ in CeCoIn$_5$/YbCoIn$_5$ superlattices. (a) $H_{c2}(\theta)$ for the S$^{\ast}$-type (5:8:5:2) superlattice. (b)  Comparison of $H_{c2}(\theta)$ in   (5:8:5:2) (red triangles, left axis),  (5:7:5:3) (green circles, right axis) and  (5:5:5:5) (blue squares, right axis) superlattices near $T_c$. (c)  $H_{c2}$ of these superlattices near $T_c$, replotted in an appropriate dimensionless form. The solid lines are the fits to the data using the model described in Eq.\;(1). (d)   $\beta_{T}/\beta_{P}$, which  quantifies  the relative importance of orbital and Pauli pair-breaking effects, is plotted as a function of the thickness modulation of YbCoIn$_5$ layers $|m-m'|$ (right panel). For comparison, $\beta_{T}/\beta_{P}$ in S-type superlattice CeCoIn$_5$($n$)/YbCoIn$_5$(5) is plotted as a function of $n$ (left panel).}
\end{figure}

It has been pointed out that the ``local" ISB at interfaces, which results in the Rashba spin-orbit splitting of the Fermi surface of the CeCoIn$_5$ interface layers neighboring the YbCoIn$_5$ layers (Fig.\;1(a)), also has an impact on the superconducting properties in the superlattices {\cite{Goh12,Mar12}}.  In fact, as shown in the inset of Fig.\;3,  in S-type superlattices with $n$-UCT CeCoIn$_5$ sandwiched between 5-UCT YbCoIn$_5$ layers, $H_{c2\perp}/H_{c2\perp}^{\rm orb}(0)$ is strikingly enhanced with decreasing $n$.  Moreover, $\beta_{T}/\beta_{P}$ increases with decreasing $n$ as shown in Fig.\;4(d). These results have been interpreted as the increased importance of the local ISB with decreasing $n$, as the fraction of non-centrosymmetric interface layers increases. Following a similar reasoning, the thickness modulation of YbCoIn$_5$ layers in the S$^{\ast}$-type superlattices can be seen as the introduction of an additional  ISB to the CeCoIn$_5$ block layers (see Figs.\;1(a) and(b)).

Bulk CeCoIn$_5$ hosts an abundance of fascinating superconducting properties. Indeed,  $d_{x^2-y^2}$ superconducting gap symmetry is well established {\cite{Iza01,Par08,Sto08,An10,All13,Zho13}}.  Moreover,  a possible presence of FFLO phase {\cite{Rad03,Bia03,Kak05,Mat07,Kum11}} and unusual coexistence of superconductivity and magnetic order {\cite{You07,Ken08}} at low temperature and high field have been reported. Our CeCoIn$_5$-based superlattices, in which the degree of the ISB and consequently the Rashba splitting in each Ce-block layer are controllable, thus offer the prospect of achieving even more fascinating pairing states than the bulk CeCoIn$_5$. The availability of these superlattices provides a new playground for exploring exotic superconducting states, such as a helical vortex state {\cite{Kau05}}, pair-density-wave state {\cite{Yos12}}, complex-stripe phases {\cite{Yos13}} 
, topological superconducting state {\cite{Tan09,Sat09}} and Majorana fermion excitations {\cite{Sat10}}, in strongly correlated electron systems.  It should be noted that very recently the formation of Majorana flat band has been proposed in the present S$^{\ast}$-type superlattice structure {\cite{Yuan}}.

In summary,  we have fabricated a novel type of superconducting superlattices,  in which the thickness of CeCoIn$_5$ is kept to $n$ for the entire superlattice, while the thickness of YbCoIn$_5$ alternates between $m$ and $m'$ from one block layer to the next, forming a ($n$:$m$:$n$:$m'$) superlattice structure.  Through the measurements of the temperature and angular dependencies of the upper critical field, we find a significant suppression of the Pauli pair-breaking effect in these superlattices when the inversion symmetry breaking is introduced.  The magnitude of this suppression increases with the degree of  the thickness modulation  $|m-m'|$. These results  demonstrate that the Rashba spin-orbit interaction in each CeCoIn$_5$ block layer  is largely tunable in these modulated superlattices.  Our work paves the way for obtaining novel superconducting states through the thickness modulation in the superlattices with strong spin-orbit coupling.

We acknowledge discussions with A. I. Buzdin,  S. Fujimoto, and M. Sigrist. This work was supported by KAKENHI from the Japan Society for the Promotion of Science (JSPS) and by the ``Topological Quantum Phenomena" (Nos. 25103711 and 25103713) Grant-in-Aid for Scientific Research on Innovative Areas from the Ministry of Education, Culture, Sports, Science and Technology (MEXT) of Japan.


\end{document}